\documentclass[aps,prc,a4paper,superscriptaddress,nofootinbib,twocolumn,floatfix,letter, showkeys]{revtex4-2}

\usepackage{graphicx}
\usepackage{amsmath,amssymb,bbold}
\usepackage{subfig}
\usepackage{hyperref}
\usepackage{comment}
\usepackage{amsthm,amsmath,amssymb}\usepackage{mathrsfs}
\usepackage{xcolor}
\usepackage{CJKutf8}

\begin{document}


\title{System-size dependence of $\gamma$-jet modifications in heavy-ion collisions}



\author{Yu-Xin Xiao}
\affiliation{Key Laboratory of Quark \& Lepton Physics (MOE) and Institute of Particle Physics, Central China Normal University, Wuhan 430079, China}

\author{Qing-Fei Han}
\affiliation{Key Laboratory of Quark \& Lepton Physics (MOE) and Institute of Particle Physics, Central China Normal University, Wuhan 430079, China}

\author{He-Xia Zhang}
\email{hxzhang@qztc.edu.cn}
\affiliation{Key Laboratory of Information Functional Material for Fujian Higher Education, College of Physics and Information Engineering,  Quanzhou Normal University, Quanzhou, 362000, China}

\author{Hanzhong Zhang}
\email{zhanghz@mail.ccnu.edu.cn}
\affiliation{Key Laboratory of Quark \& Lepton Physics (MOE) and Institute of Particle Physics, Central China Normal University, Wuhan 430079, China}


\date{\today}

\begin{abstract}
Medium modifications of the $\gamma$-triggered jets are investigated with the Linear Boltzmann Transport (LBT) model in heavy-ion collisions with varying system sizes, focusing on centrality dependence in Pb+Pb and Xe+Xe collisions at the LHC. Our numerical results reveal that jets produced in central collisions exhibit a wider transverse asymmetry ($A_N^y$) distribution, a broader jet shape, and a more pronounced $\gamma$-jet transverse momentum imbalance ($X_{J\gamma}=p_T^{\rm jet}/p_T^\gamma$) compared to peripheral collisions. These effects arise from the longer path length and stronger jet-medium interactions in central collisions, leading to enhanced jet quenching and medium response. Our findings demonstrate that the magnitude of $\gamma$-jet modifications is sensitive to the size and centrality of the collision system, with larger systems inducing more significant alterations due to increased energy loss and medium feedback.

\end{abstract}


\maketitle

\section{Introduction}

Over the past twenty years, the Relativistic Heavy Ion Collider (RHIC) at Brookhaven National Laboratory and the Large Hadron Collider (LHC) at CERN have executed a series of experiments on relativistic heavy-ion collisions. These experiments provide an unparalleled experimental environment to explore the properties of matter under strong interactions. One of the most notable discoveries is the formation of a new state of matter, termed the Quark Gluon Plasma (QGP). It is a deconfined phase of quarks and gluons, hypothesized to have existed within the neutron star and in the primordial universe microseconds after the Big Bang.

Jets, which are showers of energetic particles produced by the fragmentation of large transverse partons, serve as a critical tool for examining the characteristics of the QGP medium.
In relativistic heavy-ion collisions, jets or high transverse momentum ($p_T$) partons, generated in the early stage of collisions, propagate through the quark-gluon plasma (QGP), undergo multiple scattering interactions with the QGP medium, and experience energy loss. This phenomenon is known as jet quenching \cite{Gyulassy:1990ye, Wang:1992qdg, Qin:2015srf}.
This effect has been detected in experiments conducted at both the Relativistic Heavy Ion Collider (RHIC) \cite{PHENIX:2010nlr, STAR:2002ggv} and the Large Hadron Collider (LHC) \cite{ALICE:2012vgf, ATLAS:2011ah, CMS:2012tqw, CMS:2017xgk}. Furthermore, it has been noted that large $p_T$ dihadrons, gamma-hadrons, and dijets show suppression, transverse momentum balance \cite{Li:2024uzk} and azimuthal decorrelation \cite{Wang:2002ri, Vitev:2004bh, Zhang:2007ja, Zhang:2009rn, PhysRevLett.104.132001}, and the structure of the full jet is significantly altered by the QGP medium \cite{Qin:2010mn, He:2011pd, Young:2011qx, Barata:2023zqg}, as evidenced at both RHIC and LHC \cite{STAR:2006vcp, Casalderrey-Solana:2010bet, ALICE:2010khr, ATLAS:2010isq, CMS:2018jco}.

Theoretical studies have indicated that the intensity of jet quenching is directly proportional to the jet transport coefficient $\hat{q}$, defined as the averaged transverse momentum broadening squared per unit length along the trajectory of a parton \cite{Baier:1996sk, Baier:1996kr, Baier:1998kq, Guo:2000nz, Wang:2001ifa, Casalderrey-Solana:2007xns, Majumder:2009ge}. In recent years, research concerning $\hat{q}$ has gained significant attention \cite{JETSCAPE:2020mzn, JETSCAPE:2020shq, JETSCAPE:2021ehl, Han:2022zxn, Xie:2024xbn, Xie:2022ght, He:2018gks}.
The variable $\hat{q}$ has the ability to reflect numerous properties of the QGP medium. It is associated with the energy loss of the jet and is directly linked to the temperature and density of the medium. In the context of heavy-ion collisions, the distribution of $\hat{q}$ and its spatial gradient within an expanding QGP fireball is non-uniform. The temperature and density reach their peak at the center of the fireball, where the value of $\hat{q}$ also attains its maximum.
The initial spatial distributions are determined by the nuclear geometry of two colliding nuclei. The subsequent time evolution can be ascertained through a hydrodynamic model, utilizing an initial condition model commonly referenced in previous studies \cite{Miller:2007ri, Pang:2012he, Pang:2018zzo, Schenke:2010nt, Schenke:2010rr, Gale:2012rq, Ryu:2015vwa, Moreland:2014oya, Ke:2016jrd}. As partons traverse the non-uniform QGP medium, insights into the evolution can be gleaned by examining the final jets. Notably, the two-dimensional (2D) jet tomography technique, which integrates both transverse \cite{He:2020iow} and longitudinal \cite{Zhang:2007ja, Zhang:2009rn} tomography methods introduced in our prior work, facilitates the identification of the initial production position of the $\gamma$-jets \cite{Xiao:2024ffk}.

Jets triggered by gauge bosons, such as photons, are regarded as a ``golden channel" among jet quenching observables due to their distinctive properties. Since these bosons are produced in the initial hard scattering process and do not undergo strong interactions with the quark-gluon plasma (QGP) medium, they serve as an effective probe for studying jet energy loss. By analyzing boson-triggered jets, we can more reliably access the initial transverse momentum of the associated-jet information that would otherwise be obscured by medium interactions. Among the various boson-triggered jet studies, $\gamma$-tagged jets constitute one of the most prominent and well-investigated categories \cite{Anderson:2022nxb, Xie:2020zdb, Luo:2021voy, Chen:2020kex, Yang:2021qtl}.

In our previous study \cite{Xiao:2024ffk}, we employed 2D jet tomography to select jet events with varying initial production positions on the transverse plane. This approach amplifies the medium modification signatures of $\gamma$-jets—including both jet quenching and jet-induced medium response—by concentrating on jets produced at the central region of the QGP medium in 0-10\% Pb+Pb collisions at $\sqrt{s_{NN}}=5.02$ TeV. Our findings reveal that jet modifications in the QGP medium are governed by two key factors:
First, they depend on the spatial distribution of the jet transport coefficient $ \hat{q} $ within the medium, which determines the jet quenching magnitude and energy loss. The energy dissipated by hard partons is redistributed from the jet core to large-angle regions relative to the jet axis, thereby enhancing the medium response.
Second, these modifications correlate with the spatial gradient of the medium density. Given the medium's inhomogeneous distribution, individual jet partons within the jet cone undergo distinct evolution processes, ultimately generating an asymmetric jet shape.

This framework offers an alternative approach to probe jet modifications by leveraging QGP fireballs of different sizes produced in collisions with varying centralities. In heavy-ion collisions, central collisions create larger, hotter QGP systems, whereas peripheral collisions yield smaller, cooler systems \cite{ATLAS:2010isq, ALICE:2010suc, Sas:2022wjp, Nijs:2020roc}. Consequently, jets traversing central collisions exhibit more pronounced quenching and medium-induced modifications compared to peripheral cases.
Moreover, system sizes differ across collision species. For example, at fixed centrality, Xe+Xe collisions generate smaller systems than Pb+Pb collisions \cite{ALICE:2018lao, ALICE:2019zfl, Zakharov:2018ctz}.
In this work, we systematically analyze $\gamma$-jet modifications across different system scales by comparing centrality-dependent jet observables in Pb+Pb and Xe+Xe collisions. We anticipate that future experimental studies of jet modification and asymmetry will advance our understanding of jet transport properties and medium response dynamics.

The remainder of this paper is organized as follows. Sec.~\uppercase\expandafter{\romannumeral2} introduces our framework, specifically the LBT model. 
Sec.~\uppercase\expandafter{\romannumeral3} discusses the medium modification of $\gamma$-jet with various centrality intervals in Pb+Pb collisions and Xe+Xe collisions, respectively.
A summary is given in Sec.~\uppercase\expandafter{\romannumeral4}.

\section{Framework}

The Linear Boltzmann Transport (LBT) model \cite{Li:2010ts, He:2015pra, Cao:2016gvr, Wang:2013cia, Cao:2020wlm, Luo:2023nsi} was developed to investigate the jet propagation within the QGP medium in heavy-ion collisions, which not only describes parton energy loss but also accounts for jet-induced medium excitation. It has been used to describe experimental data on the suppression of single hadrons \cite{Cao:2016gvr}, single jets \cite{He:2018xjv}, $\gamma$-hadron \cite{Chen:2017zte}, and $\gamma$-jet correlations \cite{Luo:2018pto} in heavy-ion collisions at both RHIC and LHC energies.

In this work, jet propagation (including shower and recoil partons), parton energy loss, and the medium response in QGP medium due to the interactions between jet and medium are simulated within the LBT model, which is described by the linear Boltzmann equations,
\begin{align}
\begin{split}
p_a & \cdot \partial f_a = \int \prod_{i=b,c,d} \frac{d^3 p_i}{2E_i (2\pi)^3} \frac{\gamma_b}{2}(f_c f_d - f_a f_b) |\mathcal{M}_{ab\rightarrow cd}|^2 \\
&\times S_2(\hat{s}, \hat{t}, \hat{u}){2\pi}^4 \delta^4 (p_a + p_b - p_c - p_d) + {\rm inelastic},
\end{split}
\end{align}
where $|\mathcal{M}_{ab\rightarrow cd}|$ is the leading-order elastic scattering amplitude that is defined by the Mandelstam variables $\hat{s}, \hat{t}$, and $\hat{u}$. $f_i=(2\pi)^3 \delta^3 (\vec{p}-\vec{p}_i)\delta^3 (\vec{x}-\vec{x_i}-\vec{v_i}t)$ is the phase space density for jet partons before ($i=a$) and after ($i=c$) scattering.
$f_i=1/(e^{p_i\cdot u/T}\pm 1)$ is the phase space distribution of thermal partons before ($i=b$) and after ($i=d$) scattering in the QGP medium with local temperature $T$ and fluid 4-velocity $u$. The regularization factor,
\begin{equation}
S_2(\hat{s}, \hat{t}, \hat{u}) = \theta (\hat{s}\geq 2\mu_D^2)\theta(-\hat{s}+\mu_D^2\leq\hat{t}\leq -\mu_D^2),
\end{equation}
here is used to avoid the collinear divergence in $|\mathcal{M}_{ab\rightarrow cd}|$, and the Debye screening mass is defined as $\mu_D^2=3g^2T^2/2$.

In the LBT model, the description of inelastic processes of medium-induced gluon radiation is based on the high-twist approach \cite{Guo:2000nz, Zhang:2003wk} with the radiative gluon spectrum,
\begin{equation}
\frac{dN^a_g}{dzdk^2_{\perp}d\tau} = \frac{6\alpha_s P_a(z)k^4_\perp}{\pi (k^2_\perp + z^2 m^2 )^4} \frac{p\cdot u}{p_0} \hat{q}_a (x) \sin^2 \frac{\tau - \tau_i}{2\tau_f},
\end{equation}
where $m$ is the mass of the propagating parton $a$, $z$ and $k_\perp$ are the radiated gluon's energy fraction and transverse momentum, respectively.
The strong coupling constant is denoted as $\alpha_s$, and $\hat{q}_a$ represents the jet transport coefficient, which is defined as the squared transverse momentum transfer per unit path length for the parton $a$ traversing through the medium.
$P_a(z)$ is the splitting function for $a\rightarrow a+g$, and $\tau_f = 2Ez(1-z)/(k^2_\perp + z^2 M^2)$ represents the formation time of the radiated gluon \cite{Luo:2018pto}. 
In the LBT model, both the jet shower partons and the recoiled medium partons follow the Boltzmann equation for their transport. After being produced, they undergo further elastic and inelastic scattering processes. 
The depletion of the medium due to back-reaction is also taken into account through ``negative" partons whose energy and momentum must be subtracted from the final observables.
The recoiled partons and the ``negative" partons are referred to as the medium-response in our description.

As a fundamental component of our jet propagation framework, the Linear Boltzmann Transport (LBT) model describes jet evolution within the quark-gluon plasma (QGP) medium. Our comprehensive simulation integrates multiple theoretical components: the TRENTo initial condition model, (3+1)-dimensional viscous hydrodynamics (CLVisc), PYTHIA8 event generator, and FastJet package. The initial parton shower for $\gamma$-jet events is generated using PYTHIA 8 \cite{Sjostrand:2006za, Sjostrand:2007gs}. The TRENTo model \cite{Moreland:2014oya, Ke:2016jrd} provides both the initial energy density distribution and collision geometry, including binary collision counts ($N_{\rm coll}$) that determine $\gamma$-jet production positions. These energy density distributions serve as initial conditions for the CLVisc (3+1)-D viscous hydrodynamic evolution \cite{Pang:2012he, Pang:2018zzo} of the QGP medium. The LBT model simulates jet parton propagation through the evolving QGP medium, with scattering rates calculated using local temperature and fluid four-velocity from CLVisc at each timestep. The hydrodynamic simulation uses initial time $\tau_0=0.6$ fm/c and freeze-out temperature $T_f=137$ MeV, defining the duration of parton-medium interactions. Final jet reconstruction employs the anti-$k_t$ algorithm implemented in FastJet \citep{Cacciari:2011ma, Cacciari:2005hq}.

\begin{figure}[htbp]
\centering
\includegraphics[width=0.35\textwidth]{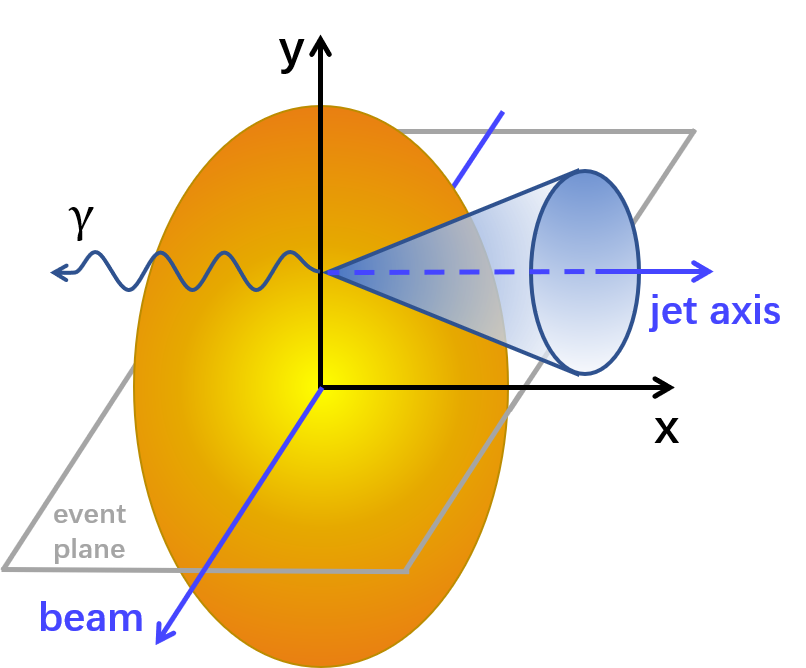}
\captionsetup{justification=raggedright}
\caption{(Color online) Illustration of the $\gamma$-jet configuration relative to the event plane.}
\label{illustration}
\end{figure}

In the following calculations, we select the trigger photon with $p_T^\gamma>$ 60 GeV/$c$ of $\gamma$-jet events in Pb+Pb collisions at $\sqrt{s_{NN}} = 5.02$ TeV and Xe+Xe collisions at $\sqrt{s_{NN}} = 5.44$ TeV. The cone size of the $\gamma$-triggered jets is set as $R=0.3$, and the lower threshold of transverse momentum for the jets is set at $p_T^{\rm jet} > 30$ GeV/c, with associated partons selected with $p_T^{\rm assoc}>1$ GeV/$c$. The pseudorapidities for the $\gamma$ and jets are constrained within $|\eta_{\gamma}|<1.44$ and $|\eta_{\rm jet}|<1.6$, respectively, and their azimuthal angle differences $|\Delta\phi_{j\gamma}|$ are restricted to be larger than $(7/8)\pi$.
Shown in Fig.~\ref{illustration} is the illustration of the $\gamma$-jet configuration relative to the event plane.
In all our calculations for Pb+Pb and Xe+Xe collisions, we assume the event plane is oriented along the $x$-axis. The wavy arrow pointing toward the negative $x$-direction denotes the selected trigger photon direction, which lies parallel to the event plane. Correspondingly, the associated jet is approximately aligned along the positive $x$-axis, as illustrated by the cone in Fig.~\ref{illustration}.
On one hand, unifying the initial directions of the jets can prevent the values of $A_N^{\vec{n}}$ from being averaged out statistically, which is crucial for the calculation of $A_N^{\vec{n}}$. On the other hand, constraining the initial directions of the jets helps us better regulate the paths of jet propagation through the QGP. This enables us to study the jet quenching effect relying on the propagation path more directly.

\section{Jet tomography within different system size via Pb+Pb and Xe+Xe collisions}

\subsection{Jet modification of $\gamma$-jet in Pb+Pb collisions}

Due to the non-uniform QGP medium, an asymmetry of jet denoted as $A_N^{\vec{n}}$ caused by the gradient of the medium density (or the $\hat{q}$ gradient) can be introduced to localize the initial position of jets production perpendicular to the propagation direction, when it propagates through the QGP medium \cite{He:2020iow}. The asymmetry $A_N^{\vec{n}}$ used for transverse jet tomography is defined as,
\begin{equation}
A_N^{\vec{n}} = \frac{\sum_a\int d^3 r d^3 k f_a\left( \vec{k}, \vec{r} \right) {\rm Sign} \left( \vec{k} \cdot \vec{n} \right)}{\sum_a\int d^3 r d^3 k f_a\left( \vec{k}, \vec{r} \right)},
\label{eqn:asym}
\end{equation}
where $f_a$ is the phase-space distribution of partons, $\vec{n}$ denotes the normal direction of the plane which is defined by the beam direction and the direction of the trigger particle, as the positive $y$-axis in our set-up shown in Fig.~\ref{illustration}. We sum up all the partons inside a jet.

\begin{figure}
\centering
\includegraphics[width=0.49\textwidth]{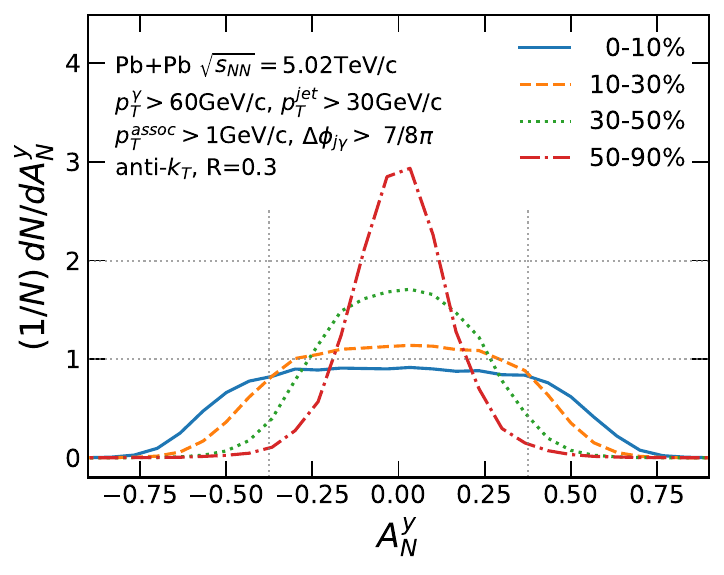}
\captionsetup{justification=raggedright}
\caption{(Color online)  The $\gamma$-jet production rates $(1/N) dN/dA_N^{y}$ as a function of $A_N^{y}$, with four centrality intervals, specifically 0-10\%, 10-30\%, 30-50\%, and 50-90\%, in Pb+Pb collisions at $\sqrt{s_{NN}}=5.02$ TeV. 
}
\label{asy_dist_pbpb}
\end{figure}

\begin{figure}
\centering
\includegraphics[width=0.49\textwidth]{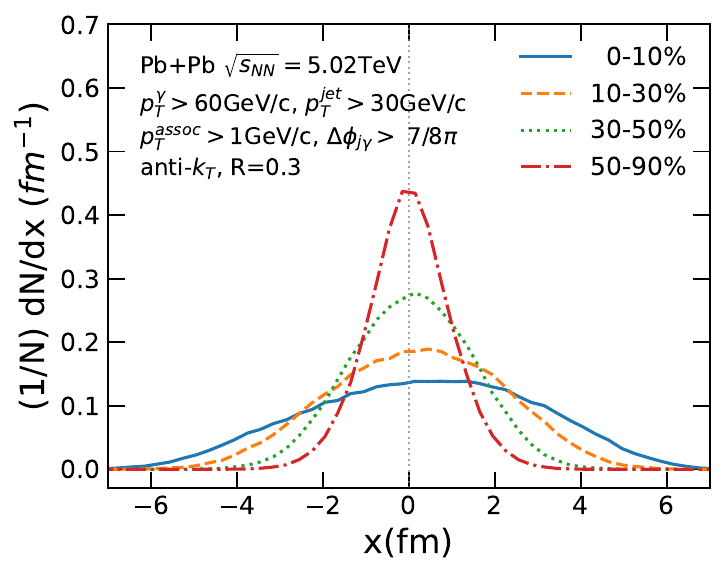}
\captionsetup{justification=raggedright}
\caption{(Color online)  The $\gamma$-jet production rates $(1/N) dN/dx$ as a function of initial jet production position $x$, with four centrality intervals, specifically 0-10\%, 10-30\%, 30-50\%, and 50-90\%, in Pb+Pb collisions at $\sqrt{s_{NN}}=5.02$ TeV.}
\label{y_dist_pbpb}
\end{figure}

We first examined the transverse asymmetry ($A_N^{y}$) distribution in Pb+Pb collisions across different centrality intervals. As shown in Fig.  \ref{asy_dist_pbpb}, our numerical results demonstrate the $\gamma$-jet production rates as functions of $A_N^{y}$ for four centrality classes (0-10\%, 10-30\%, 30-50\%, and 50-90\%) in Pb+Pb collisions at $\sqrt{s_{NN}}=5.02$ TeV. The $A_N^{y}$ distribution exhibits greater broadening in central collisions compared to peripheral collisions. As the collision system transitions from central to peripheral configurations, the $A_N^{y}$ distribution progressively narrows. This observation suggests that jets propagating parallel to the event plane show higher transverse asymmetry in central collisions than in peripheral collisions.
In central collisions, jets traversing the QGP experience longer propagation paths, which lead to more pronounced jet quenching effects.
These mechanisms lead to substantial medium-induced jet modifications, consequently producing more significant jet asymmetry.

Shown in Fig. \ref{y_dist_pbpb} are the $\gamma$-jet production rates as a function of initial jet production position $x$ with different centrality intervals, specifically 0-10\%, 10-30\%, 30-50\%, and 50-90\%, in Pb+Pb collisions at $\sqrt{s_{NN}}=5.02$ TeV. 
The results clearly demonstrate the variation in system size along the 
$x$-axis between central and peripheral collisions. Central collisions exhibit a broader initial jet production position distribution, corresponding to a larger collision system size, while peripheral collisions show a narrower distribution, indicative of a smaller system size.
Furthermore, the enhanced jet quenching effect in central collisions leads to more pronounced surface emission of jets, resulting in an asymmetric initial jet production position distribution with a distinct shift toward the positive 
$x$-axis direction (Fig. \ref{y_dist_pbpb}). As the collision centrality decreases (from central to peripheral), the jet quenching strength weakens, causing the 
$x$-position distributions to gradually regain symmetry and become narrower.
This is consistent with the conclusions from our previous work on longitudinal jet tomography \cite{Zhang:2007ja, Zhang:2009rn, Xiao:2024ffk}.

\begin{figure}
\centering
\includegraphics[width=0.49\textwidth]{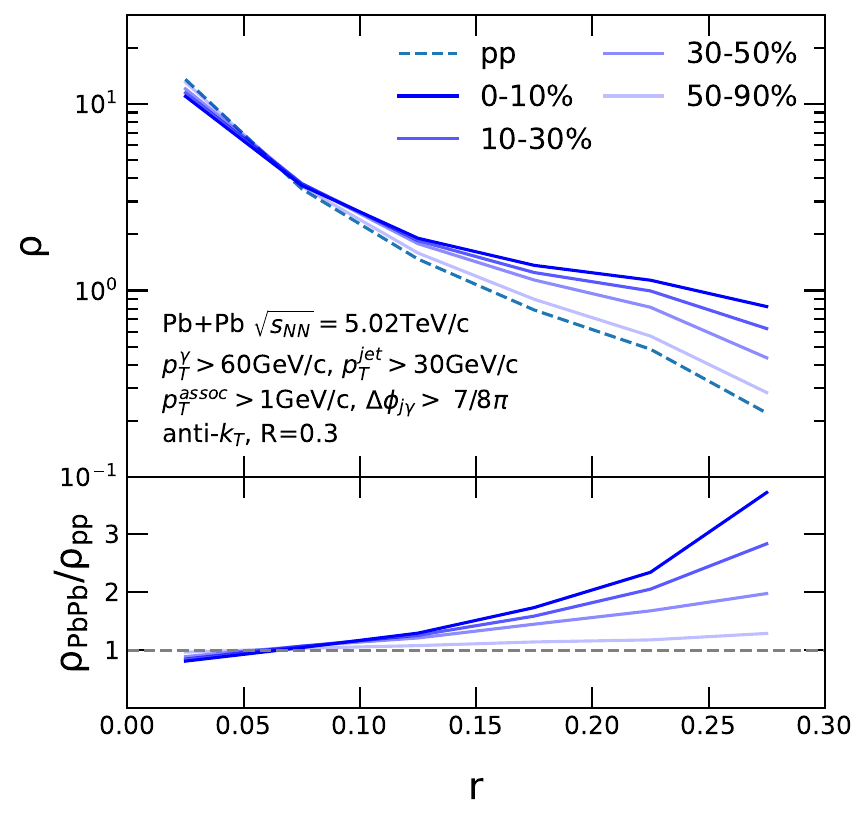}
\captionsetup{justification=raggedright}
\caption{(Color online)  Upper panel: the jet shape as a function of $r$, from different centrality intervals: 0-10\%, 10-30\%, 30-50\%, and 50-90\%, in Pb+Pb and p+p collisions at $\sqrt{s_{NN}}=5.02$ TeV, respectively.
Lower panel: The ratio of the jet shapes in Pb+Pb collisions (for each centrality interval) to the jet shape in p+p collisions.}
\label{jetshape_pbpb}
\end{figure}
 
The jet shape, which characterizes the transverse momentum distribution of particles within the jet cone relative to the jet axis, serves as a crucial observable for investigating jet transport properties and medium response. It is defined as the probability density function of transverse momentum distribution inside the jet cone\cite{CMS:2018jco, Luo:2018pto}, 
\begin{equation}
    \rho(r) = \frac{1}{\Delta r} \frac{1}{N_{\rm jet}}\sum_{\rm jet}\frac{\sum_i p_T^i (r-\frac{1}{2}\Delta r, r+\frac{1}{2}\Delta r)}{\sum_i p_T^i(0, R)},
    \label{eqn:shape}
\end{equation}
measured from the parton distribution within the jet cone of the reconstructed jets,
where $r = \sqrt{\left( \eta - \eta_{\rm jet} \right)^2 + \left( \phi - \phi_{\rm jet} \right)^2}$ denotes the distance between the associate partons and the jet axis in the plane defined by pseudorapidity $\eta$ and azimuthal angle $\phi$.
The total energy from the $i$-th jet within the circular annulus, defined by an inner radius $r_1=r - \Delta r / 2$ and outer radius $r_2=r + \Delta r / 2$, is given by $p_T^i(r_1, r_2)=\sum_{{\rm assoc}\in \Delta r}p_T^{\rm assoc}$, where $\Delta r=r_2 - r_1$ represents the width of the annulus.
Additionally, the total energy inside the jet cone with a radius $R$ for the $i$-th jet is denoted by $p_T^i(0, R)$. The final result sums the total number of jets $N_{\rm jet}$ analyzed.

Our previous study on 2D jet tomography has demonstrated that the observed jet shape broadening is correlated with the selection of jets exhibiting larger$A_N^{y}$ values and lower $p_T^{\rm jet}$ in 0-10\% $Pb+Pb$ collisions \cite{Xiao:2024ffk}. This observation indicates that jets originating from the central region of the QGP undergo the most significant medium-induced modifications, which can be explained by their extended propagation paths through the dense medium and the stronger jet quenching effects occurring at the higher temperatures characteristic of the QGP core region.

This study calculates the jet shapes corresponding to different centrality intervals. The upper panel of Fig. \ref{jetshape_pbpb} presents the dependence of $\gamma$-jet shapes on the radius $r$ in Pb+Pb collisions at $\sqrt{s_{NN}}=5.02$ TeV across various centrality intervals (specifically 0–10\%, 10–30\%, 30–50\%, and 50–90\%), with comparisons made to jet shapes in p+p collisions. The lower panel shows the ratio of jet shapes between Pb+Pb and p+p collisions.
Experimental observations reveal that in both Pb+Pb and p+p collisions, all jet shapes exhibit a monotonic decrease with increasing radius $r$, indicating that the energy distribution within the jet cone is predominantly concentrated in the core region near the jet axis. By comparing jet shapes in p+p and Pb+Pb collisions, we find that across different centrality ranges, the jet shapes in Pb+Pb collisions are generally broader than those in p+p collisions. Moreover, within Pb+Pb collisions, the jet shapes in central collisions are broader than those in peripheral collisions.
Numerical results in Fig. \ref{jetshape_pbpb} demonstrate that in Pb+Pb collisions, compared to p+p collisions, the transverse momentum lost by hard partons in the jet core region is partially transferred to soft medium-response particles at large angles relative to the jet axis. Furthermore, the differences in jet shapes among different centrality intervals suggest that jets originating from larger collision systems (e.g., central collisions) undergo more significant medium modification effects due to longer propagation paths, stronger jet quenching, and more prolonged evolution times.

\begin{figure}[htbp]
\centering
\includegraphics[width=0.49\textwidth]{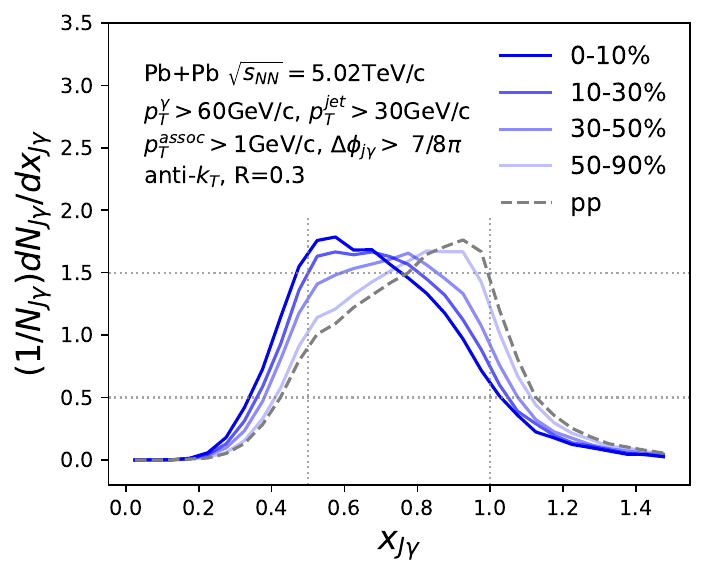}
\captionsetup{justification=raggedright}
\caption{(Color online)  The $X_{J\gamma}$ distribution of $\gamma$-jets with four centrality intervals: 0-10\%, 10-30\%, 30-50\%, and 50-90\%, in Pb+Pb and p+p collisions at $\sqrt{s_{NN}}=5.02$ TeV, respectively.
}
\label{xj_dist_test}
\end{figure}

Jets produced in different centrality intervals demonstrate distinct medium modification effects, as evidenced by the transverse momentum imbalance between $p_T^{\rm jet}$ and $p_T^{\rm \gamma}$. The $\gamma$-jet  transverse momentum imbalance is quantified by the ratio $X_{J\gamma}=p_T^{\rm jet}/p_T^{\gamma}$. Fig. \ref{xj_dist_test} presents the $\gamma$-jet production rate as a function of $X_{J\gamma}$, showing results for four centrality intervals (0-10\%, 10-30\%, 30-50\%, and 50-90\%) in Pb+Pb collisions at $\sqrt{s_{NN}}=5.02$ TeV, with p+p collision results (dashed curve) provided for reference. Our calculations reveal that compared to p+p collisions, the $X_{J\gamma}$ distributions for all four centrality intervals shift toward lower values, indicating jet quenching effects. Notably, jets from central collisions exhibit a more significant leftward shift in $X_{J\gamma}$ distributions than those from peripheral collisions, reflecting stronger quenching strength and medium modification effects in larger collision systems, which are consistent with previous discussions.

\subsection{Jet modification of $\gamma$-Jet in Xe+Xe collisions}

Furthermore, while comparing the QGP size differences between central and peripheral Pb+Pb collisions, we note that different nucleus-nucleus collision systems also demonstrate significant variations in system size. In this study, we have extended our investigation to include jet modifications in Xe+Xe collisions - a system characterized by smaller dimensions compared to Pb+Pb collisions.

\begin{figure}[htbp]
\centering
\includegraphics[width=0.49\textwidth]{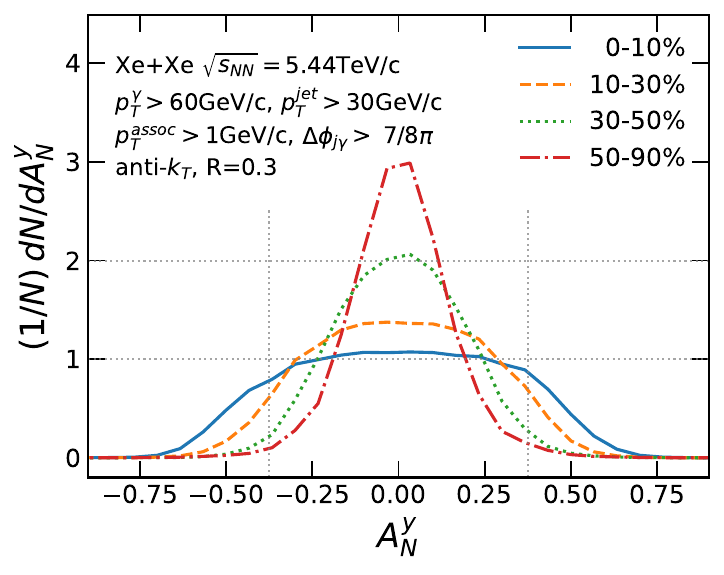}
\captionsetup{justification=raggedright}
\caption{(Color online)  The $\gamma$-jet production rates $(1/N) dN/dA_N^{y}$ as a function of $A_N^{y}$, with four centrality intervals: 0-10\%, 10-30\%, 30-50\%, and 50-90\%, in Xe+Xe and p+p collisions at $\sqrt{s_{NN}}=5.44$ TeV, respectively.}
\label{asy_dist_xexe}
\end{figure}

\begin{figure}[htbp]
\centering
\includegraphics[width=0.49\textwidth]{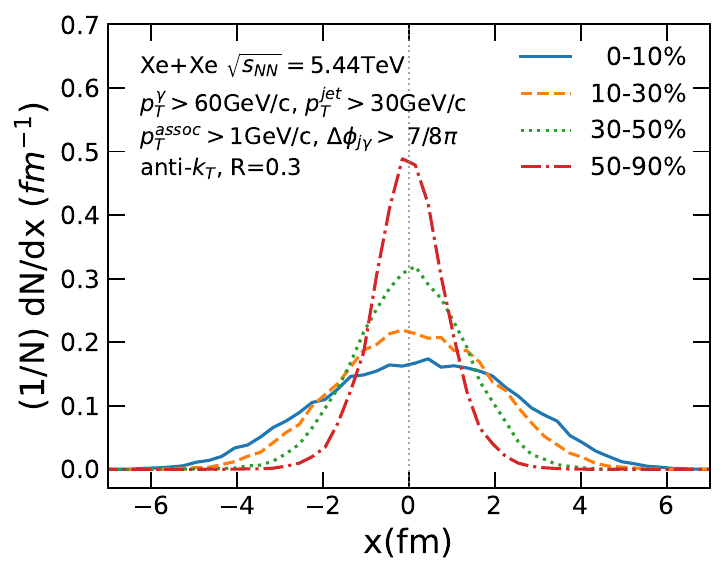}
\captionsetup{justification=raggedright}
\caption{(Color online)  The $\gamma$-jet production rates $(1/N) dN/dA_N^{y}$ as a function of initial jet production position $x$, with four centrality intervals, specifically 0-10\%, 10-30\%, 30-50\%, and 50-90\%, in Xe+Xe collisions at $\sqrt{s_{NN}}=5.44$ TeV.}
\label{y_dist_xexe}
\end{figure}

Shown in Fig. \ref{asy_dist_xexe} are the $\gamma$-jet production rates as a function of $A_N^{y}$ for different centrality intervals, specifically 0-10\%, 10-30\%, 30-50\%, and 50-90\%, in Xe+Xe collisions at $\sqrt{s_{NN}}=5.44$ TeV.
Similar trends are observed in Xe+Xe collisions, where the 
$A_N^y$ distribution demonstrates a broader profile in central collisions compared to peripheral collisions. This observation indicates an enhanced transverse asymmetry of jets propagating parallel to the event plane, consistent with findings in Pb+Pb collisions.
Moreover, due to the smaller system size in Xe+Xe collisions, the jet quenching strength and medium modification effects are weaker than in Pb+Pb collisions. Consequently, the 
$A_N^y$ distribution is narrower in Xe+Xe collisions than in Pb+Pb collisions at the same centrality interval, particularly in central collisions.
Similarly, the initial jet production position 
$x$ exhibits a wider distribution in central Xe+Xe collisions than in peripheral Xe+Xe collisions, as shown in Fig. \ref{y_dist_xexe}. Conversely, when comparing Xe+Xe and Pb+Pb collisions at the same centrality, the 
$x$ distribution is broader in Pb+Pb collisions than in Xe+Xe collisions.

\begin{figure}[bp]
\centering
\includegraphics[width=0.49\textwidth]{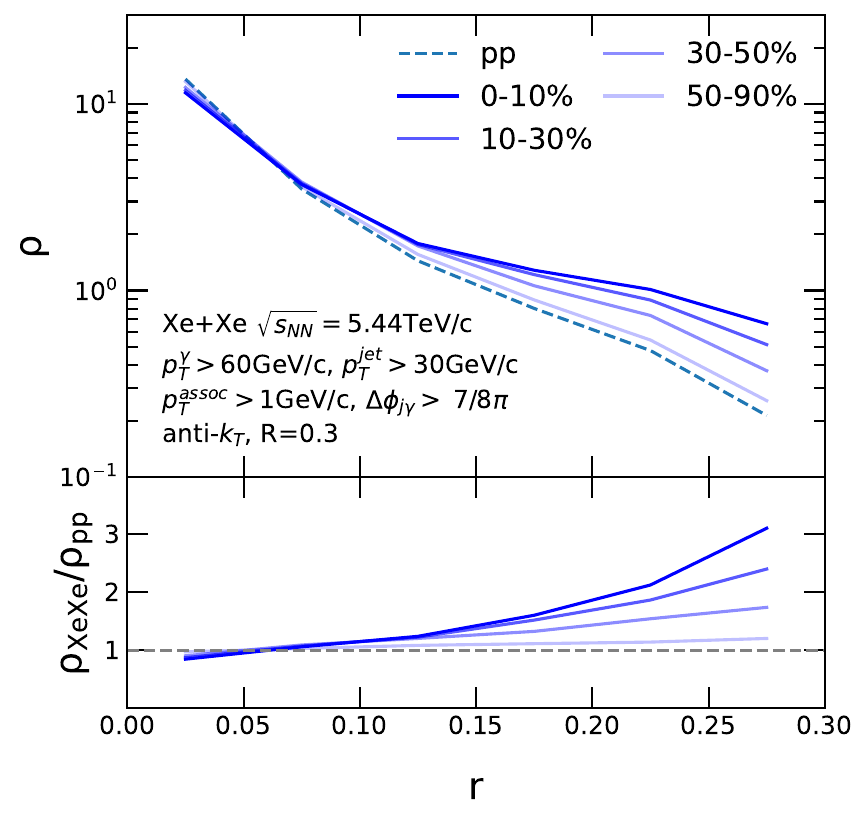}
\captionsetup{justification=raggedright}
\caption{(Color online)  Upper panel: the jet shape as a function of $r$, from different centrality intervals: 0-10\%, 10-30\%, 30-50\%, and 50-90\%, in Xe+Xe and p+p collisions at $\sqrt{s_{NN}}=5.44$ TeV, respectively.
Lower panel: The ratio of the jet shapes in Xe+Xe collisions (for each centrality interval) to the jet shape in p+p collisions.}
\label{jetshape_xexe}
\end{figure}

We also calculate the jet shapes of jets from different centrality intervals of Xe+Xe collisions. Shown in the upper panel of Fig. \ref{jetshape_xexe} are the jet shapes as a function of the radius $r$, for $\gamma$-jets produced in different centrality intervals, specifically 0-10\%, 10-30\%, 30-50\%, and 50-90\% in Xe+Xe collisions at $\sqrt{s_{NN}}=5.44$ TeV, as compared to that in p+p collisions. The lower panel is the ratio of jet shapes between Xe+Xe and p+p collisions. 
Similar to Pb+Pb collisions, Xe+Xe collisions also exhibit centrality-dependent modifications in jet shapes, with more pronounced broadening observed in central collisions compared to peripheral ones. This observation indicates a gradual transition from strong to weak jet quenching effects and medium response as the collision centrality decreases.
A systematic comparison between Pb+Pb and Xe+Xe systems further demonstrates that, at equivalent centrality intervals, jets in Pb+Pb collisions display broader shapes than their Xe+Xe counterparts. This distinction primarily stems from the larger system size in Pb+Pb collisions, which leads to stronger medium-induced modifications of jet properties relative to Xe+Xe collisions.

\begin{figure}[tbp]
\centering
\includegraphics[width=0.49\textwidth]{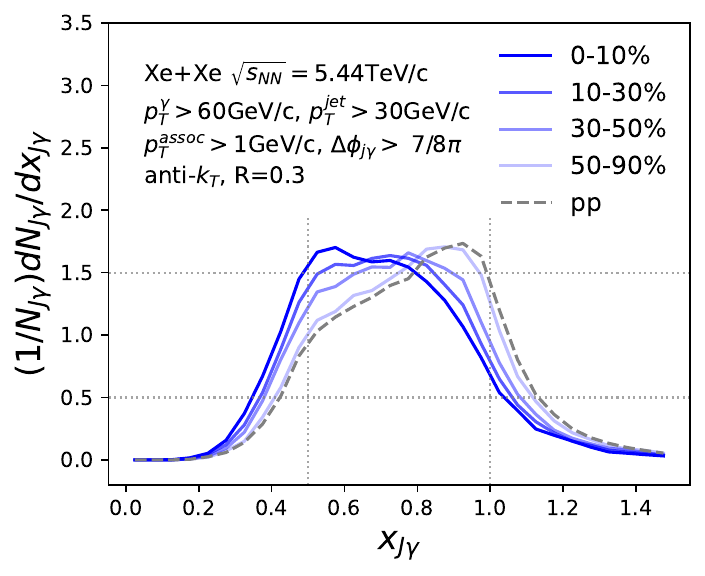}
\captionsetup{justification=raggedright}
\caption{(Color online)  The $X_{J\gamma}$ distribution of $\gamma$-jets with four centrality intervals: 0-10\%, 10-30\%, 30-50\%, and 50-90\%, in Xe+Xe and p+p collisions at $\sqrt{s_{NN}}=5.44$ TeV, respectively.}
\label{xj_dist_xexe}
\end{figure}

Finally, we examine the $\gamma$-jet  transverse momentum imbalance $X_{J\gamma}$ in Xe+Xe collisions across different centrality intervals, as shown in Fig. \ref{xj_dist_xexe}. Our calculations demonstrate that central collisions exhibit the most significant shift of the $X_{J\gamma}$ distribution toward lower values compared to p+p collisions, which reflects the strongest jet quenching effect.
The magnitude of this shift diminishes in peripheral Xe+Xe collisions, consistent with reduced quenching strength in smaller collision systems.
Notably, within equivalent centrality intervals, the $X_{J\gamma}$ distribution in Xe+Xe collisions shows less pronounced leftward shifting than observed in Pb+Pb collisions. This systematic difference indicates comparatively weaker jet quenching in Xe+Xe systems, which is in agreement with their smaller geometrical size relative to Pb+Pb collisions.
We note that, in this section, apart from the results presented in Fig. \ref{y_dist_pbpb} and Fig. \ref{y_dist_xexe}, all other results can, in principle, serve as experimental observables.

\section{Summary}

In this study, we have examined the medium modifications of $\gamma$-jets across various system sizes, particularly in different centrality intervals during Pb+Pb collisions at $\sqrt{s_{NN}}=5.02$ TeV and Xe+Xe collisions at $\sqrt{s_{NN}}=5.44$ TeV.
The LBT model has been utilized to simulate the propagation of jet partons within the QGP medium. Concurrently, the CLVisc hydrodynamic model, in conjunction with the TRENTo initial condition model, has been employed to simulate the evolution of the QGP medium across varying centrality intervals in both Pb+Pb and Xe+Xe collisions.
We have conducted a comparative analysis of the medium modifications of $\gamma$-jets originating from collision systems of diverse sizes. These size variations are contingent upon distinct centrality intervals and the nature of the colliding nucleus, whether it is Pb or Xe.
Our numerical findings indicate that, in nucleus-nucleus collisions, jets emanating from central collisions—associated with larger collision systems—demonstrate markedly enhanced medium modifications and increased jet quenching strength relative to those originating from peripheral collisions—typical of smaller collision systems.
These findings not only lead to a broader distribution of the jet transverse asymmetry, denoted as $A_N^y$, and the initial jet production position, represented by $x$, but also result in a more expanded jet shape.
Moreover, for jets originating from central collisions, there is a distinct shift towards smaller values of $X_{J\gamma}$ in the distribution of $\gamma$-jet transverse momentum imbalance, also represented by $X_{J\gamma}$.
Within the same centrality interval, Pb+Pb collisions display a larger system size than Xe+Xe collisions, resulting in more pronounced jet quenching strength and medium modification effects. As a result, jets in Pb+Pb collisions within this interval show wider distributions of both the jet transverse asymmetry, denoted as $A_N^y$, and the initial jet production position, represented by $x$. Additionally, they exhibit a broader jet shape and a more imbalanced distribution of $X_{J\gamma}$ compared to Xe+Xe collisions.
We anticipate that future experiments investigating the medium modification and asymmetric features of jets will provide valuable insights into the properties of jet transport and the responses of the medium.

\section*{Acknowledgments}

{The authors would like to thank Xin-Nian Wang for stimulating discussions. 
We thank Liqiang Zhu for providing the CLVisc data.
This work is supported by National Natural Science Foundation of China under Grants Nos. 11935007, by Guangdong Basic and Applied Basic Research Foundation No. 2021A1515110817, by Guangdong Major Project of Basic and Applied Basic Research No. 2020B0301030008, Science and Technology Program of Guangzhou No. 2019050001.
}

\bibliography{citation}

\end{document}